# Composite Variable Strategies for Continuous Endpoints

**Authors: James Bell[1], Oliver N. Keene[2], Christian Bressen Pipper[3,4], Thomas Drury[5], Lorenzo Guizzaro[6], Marian Mitroiu[7], Tobias Mütze[8], Khadija Rerhou Rantell[9], Marcel Wolbers[10], David Wright[11]**

**Affiliations**

[1] Biostatistics and Clinical Data Sciences, Elderbrook Solutions GmbH, Buckinghamshire, UK

[2] KeeneONStatistics, Maidenhead, UK

[3] Biostatistics Methods, Novo Nordisk A/S, Soeborg, Denmark

[4] Department of Public Health, University of Southern Denmark, Odense, Denmark

[5] Quantitative Science Innovation, GSK, London, UK

[6] European Medicines Agency, Amsterdam, the Netherlands.

[7] Biostatistics, Statistical Sciences and Evidence Generation, Biogen International GmbH, Baar, Switzerland

[8] Statistical Methodology, Novartis Pharma AG, Basel, Switzerland

[9] Medicines and Healthcare products Regulatory Agency (MHRA), UK

[10] Clinical Statistics and Analytics, Bayer BCC AG, Basel, Switzerland

[11] Respiratory and Immunology Biometrics and Statistical Innovation, BioPharmaceuticals R&D, AstraZeneca, Cambridge, UK

## Abstract

Composite endpoints are common in clinical trials for binary, time-to-event or recurrent outcomes. Recently, there has been interest in estimands for continuous endpoints that employ a composite variable strategy for handling intercurrent events, whereby a value on the continuous scale is defined as the outcome. In this paper we describe these continuous composite endpoints where participants who experience the intercurrent event are assigned a pre-specified value on the outcome scale (which we call the *failure value*) instead of a measurement. We describe these endpoints and the difficulty of choosing the failure value when the mean is the chosen summary statistic.

The large impact of the choice of failure value on the variance of the distribution of the endpoint is quantified as a function of the probability of the event and the standardised difference of the failure value from the mean of the unaffected values. Strategies for choosing failure values are discussed, emphasising the need to balance the clinical relevance and interpretability of the estimand with the need to reliably estimate it. Two potential solutions to situations where there is no relevant and viable choice of failure value are proposed; using failure values based on distributional characteristics to generate clinical relevance and using alternative summary measures to avoid the need to select specific values.

## Introduction

In clinical research, continuous endpoints are frequently used to quantify treatment effects, with the aim of establishing efficacy, and treatment comparisons are typically based on a comparison of group mean values. However, in most clinical trials, an important challenge is the presence of intercurrent events (ICEs), defined in ICH E9 R1 as "events occurring after treatment initiation that affect either the interpretation or the existence of the measurements associated with the clinical question of interest" [1].

The ICH E9(R1) addendum describes five different strategies that may be used to clarify how the occurrences of ICEs are reflected in the clinical question of interest. One of the five approaches is the composite variable strategy (which we refer to as the composite strategy), where the occurrence of the ICE is incorporated into the endpoint definition to form a composite variable. The resulting participant-level endpoint is then based on the measurement in participants without the ICE or else on the occurrence of the ICE itself.

A composite strategy is often used when the occurrence of a particular ICE is more relevant to the effect of interest than subsequent post-ICE outcomes [2]. While most often the ICE will represent an unfavourable outcome, it may also be a favourable one. This paper focuses on the unfavourable case, but most considerations would directly apply to the favourable scenario.

The composite strategy is particularly useful for addressing terminal events such as death [3]. No subsequent values of the endpoint can exist following a terminal event and it usually represents a clinically meaningful unfavourable outcome. Other ICE strategies may not be applicable to terminal events (e.g. the treatment policy strategy, which requires the endpoint to exist post-ICE) or may not adequately reflect the clinical meaningfulness of the event. The ICH E9(R1) addendum on estimands describes an example of using the composite strategy for a trial measuring physical functioning on a continuous scale, where participants who die are assigned a value "reflecting the lack of ability to function" [1].

Another common setting for composite strategies is where alternative or additional medication to the randomised treatment is administered due to a lack of efficacy of the original treatment regimen. In these cases, although the randomized treatment was not sufficiently effective, the new medication may improve participant outcomes. The subsequent measurements may not reflect the poor effect of initially assigned treatment and consequently including the observed outcomes in the estimation is unlikely to be relevant when the effect of the originally assigned treatment is of interest. A composite strategy can reflect the negative clinical outcome, i.e. treatment failure, for participants in this situation [4].

When using a composite strategy for a continuous variable, a numerical value is required for participants experiencing the ICE to appropriately represent the treatment failure. The chosen value needs to reflect a poor outcome for the participant taking into consideration the disease setting and the nature of the ICE. For estimands defined by a difference in means as a population-level summary measure, the choice of this value can become an important part of the definition of the endpoint (and hence the estimand) and can meaningfully affect interpretation of results.

Alternative summary statistics such as the median, trimmed mean, cumulative odds ratio (from ordinal regression), or win odds have been proposed in this situation as they can avoid this need to select a specific value for participants with the intercurrent event [5, 6, 7]. However, the mean is still typically viewed as the key summary statistic of interest, with a long history of use for continuous data and familiarity for clinicians [8].

Composite endpoints, which combine multiple outcomes into a single measure, are frequently used in clinical trials and are the endpoint of choice to demonstrate efficacy and safety in regulatory guidelines in some therapeutic settings. For example, the CHMP guideline on Alzheimer's disease discusses defining a non-responder as "a patient with a certain degree of progression or who uses additional symptomatic medication", the CHMP guideline on ulcerative colitis discusses use of the composite strategy for the intercurrent event of treatment discontinuation and the CHMP guideline on bipolar disorder discusses use of the composite strategy for patients switching from one pole of the disorder to the other [9, 10, 11]. However, the summary measure typically applied has been based on a binary responder endpoint or on the time to first occurrence of one of the events rather than the mean of a continuous endpoint. The FDA guidance on amyotrophic lateral sclerosis goes in this direction by proposing a rank-based test for evaluating a composite of both survival and (continuous-valued) function [12].

There is a small body of literature describing a numerical value being chosen to reflect the poor outcome of participants who experience the intercurrent event handled by the composite strategy, although most focus on its use in estimation rather than as part of an estimand definition. Darken et al conducted simulations based on using the $95^{th}$ centile of the distribution for a recurrent event endpoint using both single and multiple imputation (MI) [13]. In a study of nasal polyps with co-primary endpoints recorded on a limited scale, the SYNAPSE investigators used approaches based on the worst score [14]. Kahan et al describe a quality-of-life endpoint in a potential cancer trial where participants are assigned a score of "0" after they died [15]. Van Lancker et al describe a potential trial in neuroscience with a longitudinal endpoint measured on a continuous disease rating scale where the worst possible value is assigned when someone dies [16]. Mallinckrodt et al discuss the use of baseline observation carried forward (BOCF) as a possible approach [17].

Perhaps reflecting a recent surge of interest in continuous composite endpoints, two recent clinical trials in idiopathic pulmonary fibrosis (IPF) have adopted differing explicit composite estimand definitions: FIBRONEER-IPF assigned a value to participants that died corresponding to the tenth centile of observed values and estimated a mean summary measure [18]. In contrast, TETON-2 assigned the 2.5$^{th}$ centile to participant that died and estimated a Hodges-Lehmann pseudomedian summary measure [19, 20].

Defining and estimating an estimand with a continuous endpoint, a mean summary, and a composite ICE strategy presents specific and potentially substantial challenges. The choice of value assigned to participants with the ICE is part of the estimand definition. Consequently, it directly affects both what is being estimated and its underlying value. Sometimes there may not exist a choice of value with a clear clinical justification.

The choice of the failure value also affects the distribution of the endpoint. If the value chosen is too far from the underlying distribution of the original endpoint (i.e. the values in patients without the composite ICE), then the outcomes for the participants experiencing these ICEs will become outliers and the endpoint distribution strongly non-Normal. For extreme failure values, the continuous endpoint can become not just bimodal, but a *de facto* binary endpoint based on whether participants had an ICE or not, with variation in the values of observed measurements becoming comparatively unimportant in estimating the treatment effect. This has important consequences for the interpretation of the observed effect size, its relationship to the research question and for study power. Given the relative lack of literature on this topic, there is therefore a pressing need to establish reasonable principles for defining continuous composite endpoints.

This paper investigates approaches for defining a numerical value to represent an unfavourable outcome due to an ICE, as well as the consequences of these choices. It is structured as follows. First, continuous composite endpoints are defined and their utility described. Next, simple formulae are presented which quantify the impact of the chosen value on the variance, together with a discussion of the implications. Potential clinical approaches to choosing the value are then proposed. The concept and utility of definitions based on the distribution of the original endpoint are then explored. Issues around estimation for continuous composites are described, followed by those for sample size and power. Alternative summary measures are then considered. Finally, the paper summarises the main issues and discusses the wider context.

## What Are Continuous Composite Endpoints?

We define a ***continuous composite endpoint*** as a continuous endpoint for which one or more intercurrent events (ICEs) are addressed using a composite strategy. This approach involves defining a specific numerical value, which we will refer to as the ***failure value***, for participants who have experienced a particular ICE for which the composite strategy is applied. Failure values are an integral part of the estimand definition for composite strategies, explicitly reflecting that the ICE is the defined (poor) outcome contributing to the treatment effect of interest.

For clarity, discussions in this paper are written assuming the perspective of choosing a failure value for a single type of ICE but they can be applied to each type of ICE where there are multiple.

Importantly, the failure value is not a missing value to be imputed, nor is it intended to estimate what might hypothetically have happened to the participant had the ICE not occurred [17]. Instead, it reflects the fact that something undesirable happened, such as a death, initiation of rescue medication, or treatment discontinuation due to lack of efficacy. Crucially, it also implies that this is more relevant to the clinical question being addressed than any subsequent measurement of the original endpoint. Consequently, the ICE and corresponding failure value are the outcome for the participant. In this way, continuous composite endpoints aim to account for both the magnitude of treatment effect on the original scale and the occurrence of clinically meaningful events.

A consequence of this is that any measurements of the continuous variable following the ICE are irrelevant, since the outcome has already occurred. The failure value will instead be assigned, independent of any available or missing outcome data. There are therefore no subsequent missing data and no need for *participant-level* uncertainty to be accounted for and translated into the variance of the estimate (e.g. by MI). In contrast, *failure-value-level* uncertainty may still occur where it is not clear what the correct choice of value is, reflecting uncertainty over what should be estimated. This may be because either no value clearly reflects the clinical severity of the ICE, or because multiple values may be of interest. This would not translate into estimation uncertainty or a higher standard error but would require careful consideration through clinical discussion and potentially supplementary analysis.

As the failure value is part of the endpoint attribute of the estimand, there are certain consequences. Firstly, it must be pre-specified as part of the trial's estimand description. Secondly, its definition must be at a population, rather than sample, level. These are discussed later in the paper. Thirdly, the failure value definition must be independent of the treatment arm as the endpoint attribute must be consistent across treatments [1, 7, 21].

Continuous composite endpoints may be useful when both the following conditions are met:

- The ICE corresponds to a clinically significant outcome of treatment (e.g. death, need for rescue therapy),

- The continuous measurement scale is considered the most suitable to quantify the treatment effect.

To illustrate a setting where a continuous composite endpoint may be suggested, consider a Phase III trial for treatment of a chronic lung condition. The agreed primary endpoint is change from baseline in Forced Vital Capacity (FVC) at 52 weeks, a continuous measure of lung function [22]. In this example, approximately 5% of participants would be expected to die before week 52, many for pulmonary-related reasons. Death represents a clinically significant failure and renders treatment policy impossible to use due to the non-existence of subsequent outcomes. One potential approach is therefore to apply a composite strategy, whereby any participant who dies before the endpoint assessment is assigned a failure value. The choice of value is however critical.

If the endpoint had been a binary dichotomisation of the measured values, such that a participant with sufficient worsening were deemed to be a 'non-response', then it would be straightforward to assign a 'non-response' to the participants that die as all we need to establish to make this decision is that death is a bad outcome. However, for a continuous endpoint it is often difficult to find a value that appropriately reflects the severity of the ICE on the scale of interest to produce a meaningful interpretation that also has acceptable statistical properties.

In the example, it might initially seem reasonable to assign an absolute FVC value of 0 mL to these participants to reflect that death is a very severe outcome. However, assigning this as the failure value and using the mean as the summary measure induces a substantial increase in variance that may prevent meaningful interpretation, as we discuss in the next section.

## Impact of Choice of Failure Value on Variance

Although estimands derive from the objectives and clinical questions of interest in a trial, it is critical that there is an "adequately reliable" estimator [1]. There should also be awareness of the consequences of the estimand for the statistical properties of its estimators.

When a continuous composite endpoint is specified and the desired summary measure is a mean, it is necessary to consider the impact of the choice of failure value on the variance of the underlying composite distribution itself (i.e. at the population level). It is often clinically desirable that the failure value represents an outlier compared to the subpopulation of participants without the ICE, since the ICEs are often very bad outcomes for participants. However, introducing outliers to a distribution will increase its variance relative to the distribution without them. As the failure value becomes increasingly extreme, the variance of the composite distribution will increase rapidly, resulting in considerable uncertainty in estimation.

The variance of continuous composite distributions will depend on the frequency of the ICE and the extremity of the failure value. A continuous composite endpoint, $Y^{\kappa}$, can be considered to be a mixture distribution of two components; the first, an arbitrary distribution with mean $\mu$ and variance $\sigma^2$ and proportion $1-p$ that represents the values for participants without the ICE and the second a point mass with proportion $p$ and the assigned failure value $\kappa$. We also define $\Delta$ as the standardised distance of $\kappa$ from $\mu$, that is: $\Delta = \frac{\kappa-\mu}{\sigma}$, i.e. the number of standard deviations that the failure value is away from the mean of the values of participants without the ICE, reflecting the extremity of the failure value.

In appendix 1, we derive that the total variance of the composite distribution is given by:

$$VAR(Y^{\kappa}) = (1-p) \cdot \{\sigma^2 + p \cdot (\mu - \kappa)^2\}. \qquad (1)$$

We also define the relative variance, $\varphi$,as the variance of the full continuous composite distribution divided by the variance of the distribution in participants without the ICE. This is given by:

$$\varphi = \frac{VAR(Y^{\kappa})}{\sigma^2} = (1-p) \cdot \{1 + p \cdot \Delta^2\}. \qquad (2)$$

Figure 1 shows $\varphi$ as a function of $\Delta$ for different values of $p$. As can be seen, the relative variance can be substantial and may exceed two when the failure value is several standard deviations away from the mean and/or the probability of an ICE is not negligible. Since sample size requirements vary proportionately with variance, a failure value that results in a doubling of variance would require twice the sample size to maintain power than one that does not affect variance. Such large increases in sample size may be operationally, ethically and/or financially unfeasible.

Without an increase in sample size to compensate for large relative variance caused by the extreme failure value, the power of a trial may be reduced to the point that it is not statistically feasible. The reliability of estimation would also be questionable, since confidence intervals would be substantially wider than otherwise, and the mean distance of point estimates from the true value would be increased, even if bias is unaffected. ICH E9(R1) states that "Where significant issues exist … to derive an adequately reliable estimate for a particular estimand, an alternative estimand, trial design and method of analysis would need to be considered."[1]. Extreme failure values that lead to large relative variance should therefore be avoided. For the range of failure values that may be reasonably accommodated within the design and analysis of a trial, the exact choice will depend primarily on clinical interpretability, which is the subject of the next section.

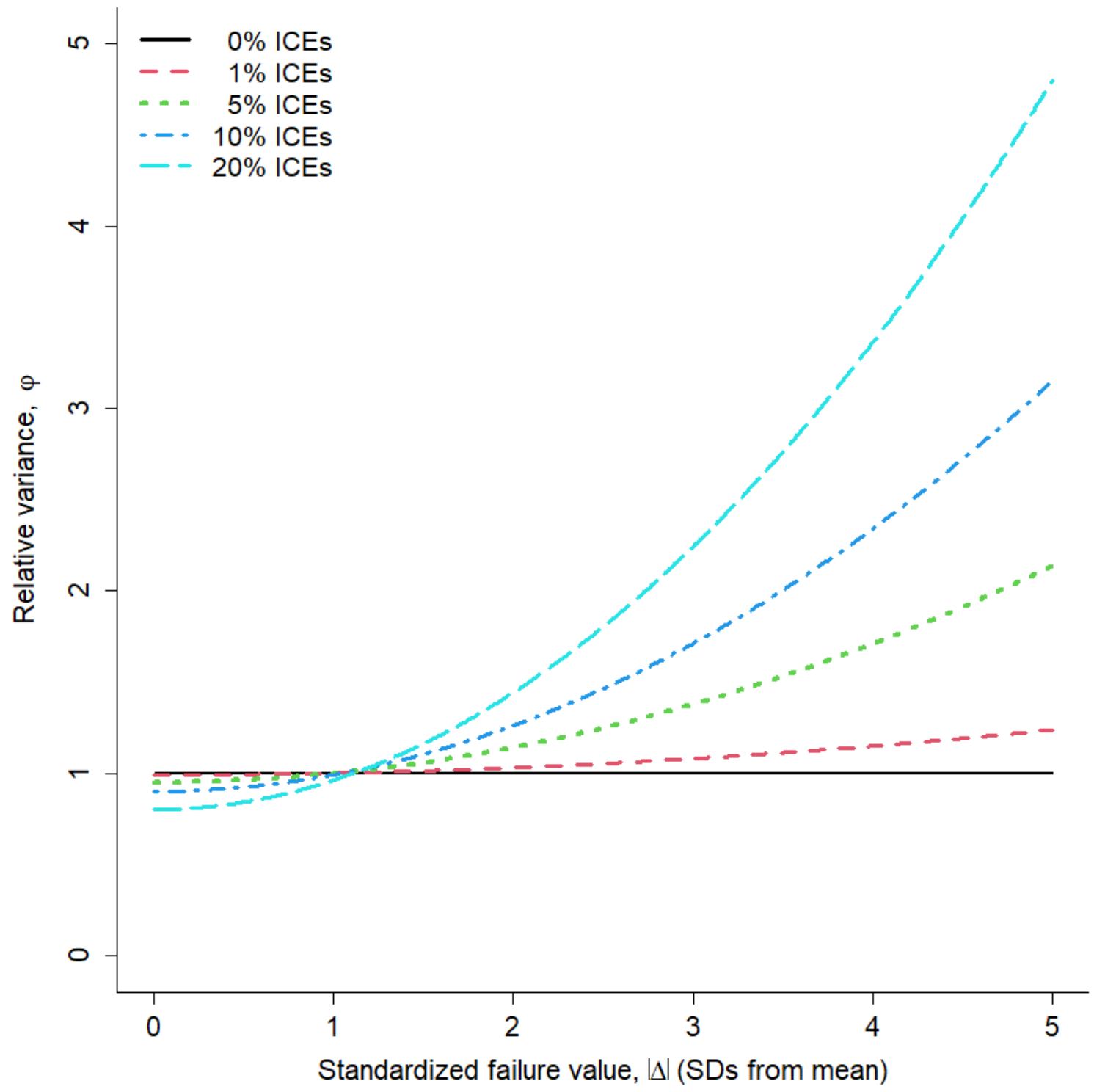

Figure 1: Visualisation of formula (2) for relative variance, $\varphi$,as a function of the standardised failure value, $\Delta$, for different frequencies of ICEs ($p$).

## Choice of Failure Values

The failure value is part of the estimand definition and its value should therefore flow primarily from the clinical question of interest that it is addressing. The above statistical concerns should act as a feasibility check (or 'veto') on choosing extreme failure values when there is a mean summary measure, but they cannot justify the selection of a particular value. The chosen failure value should be of substantial clinical meaning, reflecting the extent to which the occurrence of the ICE represents a clinical failure on the continuous scale. It can therefore be viewed as a numerical 'clinical penalty'.

Any choice is, to an extent, arbitrary and involves clinical and statistical judgement. However, the practical risk is not that judgement is used, but rather that the failure value is not accepted as relevant by stakeholders, so this should not prevent selection of a value. There are analogies with well-known situations, for example the definition of non-inferiority margins or what would constitute minimal clinically important differences [23, 24].

In some cases, there is a value that represents the worst possible outcome on the measurement scale. We refer to this as a ***hard clinical boundary***. For example, in a six-minute walk test, the worst outcome is being able to walk 0 metres [25]. In another example, for eyesight endpoints requiring the reading of letters such as best corrected visual acuity (BCVA), the worst outcome is 0 letters read [26].

Unfavourable boundary values often have some clinical relevance where the ICE is considered the worst clinical outcome possible (such as death) simply because they are the worst possible value and hence represent the worst penalty. Further clinical meaning may come where these worst values are possible to observe, since the failure value can then be directly clinically interpreted as if it had been measured. Where failure values are plausibly observable, this also implies they are not too extreme, which reduces their impact on variance. In the above examples, 0 metres in a six-minute walk test clinically represents an inability to walk. In BCVA, 0 letters represents a severe lack of eyesight. Both may be plausibly measured and have clear clinical interpretability. Consequently, they might be considered strong candidates for clinically appropriate failure values.

Unfavourable boundary values often become problematic where it is not clinically plausible for patients in the target population to obtain such values, or sometimes anything close to them. For instance, in the earlier example of FVC the unfavourable boundary value would be 0 mL, yet no living participant would, or could, have a value of zero or close to zero . Although 0 mL is highly penalising, it is physiologically implausible for living patients. This greatly reduces its clinical interpretability and simultaneously creates large outliers, whose deleterious statistical impact was previously described.

In some cases, unfavourable boundary values do not exist or are clinically implausible. An alternative when considering death might be to consider a failure value that represents a realistic clinical state for end-of-life patients; a value that participants may approach as their condition clinically deteriorates. We refer to this as a **soft clinical boundary**. Using a soft clinical boundary may have two potential advantages. First, it may be more continuous with participants' own trajectories of observed values. Second, it is likely to be worse than all, or almost all, values observed in surviving participants, so it represents a clear failure. For example, in participants with IPF, an FVC failure value for death might be pre-specified based on historical data, using a mean of FVC measurements taken within e.g. a month of participants dying of respiratory causes.

Pre-existing clinical classifications of a particular value corresponding to a bad outcome may be available in certain cases, particularly for Clinical Outcome Assessments. The advantage of using these is they build on existing clinical relevance definitions. However, such commonly accepted categorisations may not be available for the relevant endpoint. Those that are available may not be helpful either; many will just define a boundary beyond which the outcome is poor, but this may effectively only define an upper limit for the failure value.

In some cases, there will be no unfavourable boundary value or clinically accepted classifications. For instance, for concentrations of a biomarker where higher values represent worse outcomes. In such cases it may be difficult to identify a single, or even any, failure value of clinical relevance.

An alternative to using a failure value on the measured scale is to choose one that is an absolute or relative change from the baseline value. These anchor the failure values in the participant’s own baseline characteristics and may more closely reflect endpoints, which are commonly defined as changes from baseline. This is particularly relevant where it is known that a particular degree of worsening represents a failure of treatment, where there is more heterogeneity in baseline participant disease severity, or where there are no defined thresholds on the measured scale. FVC in chronic lung conditions may again be a suitable setting for this approach as there is often considerable variability in baseline lung capacity due to participant sex, age and disease severity, and it is often the decline in lung capacity over time that is being assessed.

One specific case is that of treatments of stable conditions where the intention is to improve outcomes over a certain timeframe. Here, setting a failure value to zero change from baseline may represent a meaningful clinical failure of the treatment. Consider for example, a setting of stable chronic pain where the natural history suggests there is not expected to be an underlying pattern towards greater or lesser pain over time and treatment is intended to relieve the pain. Here, a failure value of a zero change from baseline would constitute a clear failure of treatment; the level of pain has not improved. Similarly, in a weight-loss trial a zero change from baseline for weight might make sense as a failure value for the ICE of use of prohibited alternative weight-loss therapies, where it might reflect the concept that the participant was unable to lose weight on their randomised treatment.

Some cases are particularly difficult to set appropriate failure values for because the concept of penalisation does not align with clinical relevance. For example, in the case of a weight-loss trial, defining death as a weight of 0 kg is a hard clinical boundary and clearly corresponds to a clinical failure as a participant with zero weight would necessarily be dead. However, it would numerically be the greatest weight-loss possible, so give an apparent clinical benefit for the relevant treatment arm. Alternatively, defining a large positive change from baseline failure value for death would be numerically penalising and a clinical failure in a weight-loss trial. However, although weight gain eventually increases relative mortality risk, the timescale of a meaningful effect on absolute mortality is longer than the one-to-two-year span of typical participant follow-up and acute weight loss is more strongly associated with mortality [27]. Therefore, it would not seem clinically reasonable to associate death with any specific penalty value, e.g. a 10 % increase in body weight. A final option, defining death as a zero change from baseline, does not align with an intention to penalise death as a worst possible outcome but might be of interest if the intention were that death should represent a failure to lose weight.

In cases where there is no clear clinical failure value, or it is statistically impractical, clinical relevance may be derived from the distribution of the original endpoint. This is discussed in the next section.

## Defining Failure Values by Distributional Characteristics

Any possible value for the outcome variable can be regarded as a characteristic of an underlying (unknown) population distribution. For example, in a given setting, a value of 300ml may be the value of the tenth centile of the distribution of the endpoint under a control treatment. Equivalently, a characteristic of the population distribution, such as the 10th percentile, corresponds to a fixed, but unknown, value in the population. Because it is a fixed property of the population, it can be part of the definition of the estimand. The value of this characteristic of the population distribution may then be estimated from the trial.

As previously discussed, there may be situations where there is no specific numerical failure value that is both clinically and statistically appropriate. Particular population-level summary measures may however be associated with certain levels of clinical importance. For example, the 10$^{th}$ centile of the distribution or two standard deviations below the mean might be considered clinically 'bad'. This link between distributional properties and clinical importance has been used in other settings in clinical trials, such as using standardized effect sizes to assess the clinical importance of effect sizes (e.g. Cohen's d [24, 28]). Furthermore, the link between quantiles and clinical importance has also been applied to other clinical settings, e.g. for defining normal ranges for laboratory values or growth curves [29, 30].

It should be noted that properties of the sample itself should not be directly used as these cannot form the basis for a principled estimand [21]. This is because the estimand must be defined at a population-level as it is a target of estimation. An example of this is that a failure value of 'worst observed value' cannot be part of a valid estimand as it is a sample descriptor. It also does not have a corresponding population summary measure that could be used in a definition that it could be used to estimate. Other properties of the sample, such as a quantile, cannot be used directly in the definition of the failure value but can be used as estimates of corresponding summary measures from a population distribution.

Although in principle any well-defined population-level distribution could be used as the basis for the failure value, it is sensible to select one that simplifies the definition and estimation of such a summary measure. The defined distribution should be used as the basis for coding the failure value for all arms to ensure a consistent endpoint attribute and the ability to make causal inferences [7, 21].

The population-level distribution could be chosen to correspond to one or more previous trials or real-world studies, such that a value can be estimated *a priori*. It would then simply be a justification for a choice of a particular pre-specified failure value. In this case, care should be given to the potential for heterogeneity between trials leading to either failure values that are too extreme, resulting in unintended very high variance, or too mild, resulting in insufficient penalisation and unduly beneficial outcomes. Alternatively, as the remainder of this section focuses on, the population-level distribution could directly correspond to the planned study, with estimation of the failure value being performed as part of the analysis. This helps ensure complete relevance to the trial but means the value is not available during planning and conduct.

As with typical estimands, the treatment condition is important to defining the distribution from which the failure value is derived. Although theoretically it could be taken from any population, using the distribution in the population receiving the control treatment condition makes the failure value definition clearer and – to the extent that studies refer to the same population and the same control treatment condition - more transferrable between studies, since the control arm will usually most closely correspond to a standard of care.

For ease of definition and estimation, the failure value distribution should only be defined based on the population of participants without an ICE handled using the composite strategy (i.e. conditional on the non-occurrence of such ICEs). When taken from a single arm, this represents a fully-identified principal stratum that could be estimated using subsetting.

It is also necessary to consider how ICEs handled using other strategies (e.g. a treatment policy strategy) impact the distribution defining the failure value, for example, whether the tenth centile includes or excludes the effects of ICEs such as treatment discontinuation. There is no universally valid answer to this, as it may link to clinical considerations that are outside the scope of this paper, however consideration should be given to both relevance to the main analyses of interest and simplicity of estimation.

Taking the above into consideration, the failure value of distributional-based continuous composite estimands could be worded in one of the following two ways:

> *For both treatment conditions, the failure value is defined as the* ***Xth centile*** *of the distribution of the endpoint in the population of* participant*s treated with [the control] that do not experience [the event handled with the composite strategy]*
>
> *For both treatment conditions, the failure value is defined as the value that is* ***Y standard deviations below the mean*** *of the distribution of the endpoint in the population of* participant*s treated with [the control] that do not experience [the event handled with the composite strategy]*

Using a definition for a failure value based on the population’s outcome distribution potentially provides a pragmatic way to define poor outcomes in situations where there is otherwise no agreeable choice of value. However, distributional definitions do add some complexity to the estimand definition. Also, although some clinical justification may come from distributional approaches, the choice of the particular quantile or number of standard deviations from the mean is, like choice of a particular value itself, still inherently arbitrary. The corresponding failure value still represents only one point along a continuous range of responses and, by necessity, cannot capture the worst possible outcome. In any case, as the estimand is the formalisation of the clinical question of interest, the definitional distribution should always be reviewed clinically before being adopted.

## Missing Data and Estimation

When an ICE is handled with a composite strategy, the failure value is – without any uncertainty – the outcome for all participants to which the ICE occurred, so, aside from any estimation of the failure value itself when it is based on a distributional definition, there is no need for explicit or implicit imputation.

However, missing data can still arise in continuous composite settings where a composite-handled ICE has not (yet) occurred. Consequently, it is still necessary to consider missing data handling.

Notably, the failure value itself provides no useful information about missing values for participants that are known to not experience the ICE. This is because the failure value does not reflect actual measurements, but rather a pre-defined penalty regarding the existence of the ICE. It is therefore only relevant where the composite ICE has occurred. No missing data can, by definition, follow the composite ICE, so there are no such values to inform.

As failure values are often outliers, use of them in a reference set when imputing missing data for participants without the ICE can result in bias. Likewise, any observations after a composite-handled ICE will likely not be relevant (as they are affected by the composite-handled ICE) so typically will not be used to inform missing data. Missing data handling should therefore be based on relevant measured outcomes, and insertion of failure values should follow.

An important consequence is that an MMRM model without prior imputation may be inappropriate for analysing incomplete continuous composite data sets [31]. This is because MMRM would attempt to use the failure values to implicitly handle the distribution of the missing data for participants that are known not to have failed. Missing values would be treated as reflecting a weighted average of observed participant values and the failure value, depending on their proportions in the available data, which is inadequate where ICE occurrence can be assumed as always known.

For situations where the information on composite ICE occurrence (including for patients whose outcome is missing) can robustly be assumed to be complete, two potential analysis approaches are described below. They are discussed here to highlight potential directions for methodological development rather than to provide fully specified methods. The first potential approach is to perform an initial multiple imputation (MI) to handle the missing data before inserting the failure values [32]. This could be 'standard' MI with a missing-at-random assumption, or something more complicated, such as a reference-based or retrieved dropout approach, depending on the handling of other ICEs and the anticipated missing data mechanisms [33].

A second potential approach could be to analyse those participants without the failure only (e.g. by MMRM or MI by treatment arm and ANCOVA on the subset), obtain estimates from them, then perform a weighted combination with the failure value. Similar to the likelihood methods from Bell et al., weights would be based on the observed proportions of participants with the ICE and estimation would need to account for the uncertainty of these only being estimates of the true proportions [33].

Additional complexities and limitations arise when information about the occurrence of the ICE is incomplete. The frequency of this situation occurring will depend on the type of ICE and the quality of trial conduct. For example, with good conduct it should be possible to greatly limit this for the ICE of death as vital status is generally obtainable from several (including publicly available) sources. In contrast, even with good trial conduct it may be difficult, or even impossible, to identify ICEs involving usage of additional medications for participants that are lost to follow-up or have withdrawn consent. Trivial amounts of ICE-missingness might be simply handled by single imputation rules (e.g. assuming no death has occurred if no record of death is found). However, non-trivial quantities would need to be appropriately addressed during estimation, perhaps by multiply imputing ICE-occurrence jointly with the outcome, since whether failure values need to be assigned may have a large impact on both point estimates and variability. Estimation methods where there is uncertainty about the occurrence of ICEs deserve further investigation that is out of scope of this paper.

For distributional definitions, one consequence of the failure value being coded rather than explicitly defined is that there is some uncertainty in its value as the population parameter value is being estimated using a sample. This is not participant-level uncertainty since the value is consistent for all participants, but rather uncertainty about a definition. Parameter-level uncertainty should generally be smaller than uncertainty in participant values, reflecting a standard error for a parameter estimate rather than a standard deviation of values. Nonetheless, the magnitude of its impact is not well characterised and, along with methods to account for it, should be the subject of further study. The authors are therefore writing a companion paper covering this topic.

## Sample size and power

When calculating power for trials using continuous composite endpoints, simulation can be used. The simulations would need to incorporate proposed failure values and make assumptions about the mechanism of ICE occurrence including frequency, timing and links to participant values.

Alternatively, the expected impact of the failure value on the endpoint properties can be estimated analytically and incorporated into standard power calculations for continuous endpoints. This requires several considerations. Firstly, the impact of the failure value on the variance can be estimated using the formula presented above. The relative variance should be calculated separately for each arm and then either pooled or the higher value taken. For more extreme failure values, the variance, and hence power and/or sample size, will be very sensitive to deviations in assumptions. Secondly, the planned treatment effect will depend directly upon the failure value if an imbalance in composite ICEs is expected between arms. It can be estimated based on (frequency) weighted averages of the failure value and observed values in each arm. Thirdly, continuous composites are typically used in situations where either the physiological property being measured no longer exists following the ICE (e.g. following death), or the measurements are considered irrelevant (e.g. following rescue therapy). This means that more (relevant) data are typically available for composite handling that for other approaches to ICE handling, i.e. less missing data needs to be assumed.

## Alternative population summary measures

The issues around large relative variance described above arise because of the use of the difference in mean values as the estimand summary measure. An alternative to avoid the need to specify a failure value that is deemed insufficiently penalising is to change the summary measure, although this also changes the estimand and as such the clinical question of interest. One possibility (that also changes the endpoint) is to use a binary approach, i.e. to classify all participants as a success or failure on the outcome, and then to estimate e.g. an odds ratio or risk difference. In this case participants with the ICE would naturally fall into the failure category. The disadvantages of a binary approach are firstly the substantial change of interpretation and secondly the loss of information as it ignores the magnitude of the change in the endpoint [34]. Unless the proportion of participants with the ICE is particularly high, it is generally preferable to use a summary measure that preserves the continuous nature of the variable.

For continuous composite endpoints, the median is a valuable alternative to the mean as it is not sensitive to the choice of failure value if failure is ranked below all other outcomes yet it retains the link to the continuous scale. It is also likely to be somewhat familiar to clinicians, increasing its interpretability. Additionally, median-based comparisons do not rely on an assumption of residual Normality, which would be violated if many participants are assigned the same failure value at the extreme of the distribution. However, this robustness may come at the cost of reduced statistical efficiency when the endpoint distribution is close to Normal, only having an asymptotic relative efficiency (ARE) of 0.637 [35]. Furthermore, inference for differences in medians is generally less straightforward than for mean differences, often requiring resampling-based methods or additional distributional assumptions. Bootstrap methods are particularly valuable for estimating CIs when the statistic of interest has a sampling distribution that is difficult to derive analytically [36,37]. Bootstrapping allows for covariate adjustment via quantile regression, offering greater flexibility and interpretability [5]. An example of use of quantile regression is the nasal polyps trial for mepolizumab [5]. However, quantile regression has not been widely used in analysis of clinical trials and therefore this is a developing area.

An alternative approach to the median is the Hodges–Lehmann pseudomedian estimator, often used alongside the Wilcoxon rank-sum test [19]. This method assumes a constant shift in distributions between treatment groups, an assumption that the Wilcoxon test itself does not require. This assumption is not fully compatible with use of a common failure value across treatment groups.  It is estimated as the median of all pairwise differences between treatment groups (within strata, if stratified). This approach has been used in the recent TETON-2 trial [20].

The Hodges-Lehmann estimator requires a failure value to calculate the pairwise differences. However, similar to the median, once all comparisons of failure values to observed values are worse than any comparisons between observed values, the specific failure value becomes irrelevant if the proportion of failures is below the method's breakdown point of approximately 29% [35]. The advantage of the Hodges-Lehmann estimator over direct estimation of the median is it is generally much more efficient for (approximately) Normal distributions, having an ARE of 0.955, which translates to 18.5% smaller standard errors [35]. The disadvantages are that the interpretation of the pseudomedian may be more difficult than a median (corresponding to the median only for symmetrical distributions), and it is less suitable for situations expected to have high incidence of ICE(s) handled with a composite strategy as the breakdown point of 29% is lower than the median's 50%.

Other proposed summary measures include the win odds [38] and the difference in trimmed means [6]. Win odds compares the odds of a random participant in one arm having a better outcome than a random participant in another by generalised pairwise comparison between the treatments, with an adjustment for tied pairs. The trimmed mean focuses on the central portion of the endpoint distribution by excluding a specified fraction of the extremes. The disadvantages of these approaches are that the win odds loses connection with the original scale of measurement (similar to using the binary approach), while for the trimmed mean, there is no accepted rule for the amount of trimming and if the specified fraction is not large enough to cover all participants with ICEs then it may not be estimable.

## Discussion

The failure value is an attribute of the endpoint, and therefore of the estimand. The choice of failure value determines what is being estimated and its clinical interpretability must be the priority. However, when the mean is the summary statistic, this choice also affects how precisely the estimand can be estimated. Equations (1) and (2) makes this explicit. The mean treatment effect on a continuous composite endpoint combines an effect on the measurement among participants who do not experience the intercurrent event with an effect on the frequency of that event, and the failure value fixes the weight attached to each. Two trials in the same clinical setting that adopt different failure values are asking different questions, whether or not the choice was made deliberately.

Precisely because the failure value belongs to the estimand, its choice is also subject to the requirement that the resulting estimand can be estimated with adequate reliability: ICH E9(R1) is explicit that where significant issues exist in deriving an adequately reliable estimate, an alternative estimand, trial design or method of analysis should be considered. This is the main source of the tension when selecting failure values. Those that are most attractive clinically can be the most penalising, but highly penalising values also produce the largest relative variance. For the same sample size, this lowers power and yields estimates whose precision may be too poor to support a useful conclusion. Inability to estimate with acceptable precision is one way in which estimation fails to be adequately reliable, so a clinically well-motivated choice may nonetheless be unavailable.

In practice, this seems to leave four viable courses of action:

1. Raise sample size, perhaps substantially, to retain power
2. Change the failure value (i.e. to one potentially of lower clinical relevance)
3. Change the summary measure, e.g. to a median (different clinical interpretation)
4. Change the strategy for the ICE entirely (to e.g. a hypothetical one).

Raising sample size changes the design but not the estimand. However, given the potential magnitude of the relative variance, the sample size increase required may be huge, rendering some trials impractical or impossible. Therefore, the first option is likely to be only viable in cases with relatively low variance increase, i.e. where the clinically desired failure value is not too extreme. Based on the relative variance formula described in this paper, and accounting for composite-handling producing fewer missing data than under other ICE handling approaches, defining failure values up to about two standard deviations worse than the mean should be reasonably possible. In contrast, failure values beyond roughly three standard deviations worse than the mean are likely not possible to accommodate due to their substantially increased sample size requirements.

Changing the failure value to a less extreme one requires changing the estimand and hence its viability will be situational. In general, it will be necessary to discuss alternatives and the relative clinical importances with the main stakeholders. Where there is no single value of particular clinical relevance, there are likely to be a range of values that would be clinically acceptable and it is likely possible to find a value that makes the trial feasible. It should also be considered whether alternative clinical justifications for a failure value may be acceptable. One potential approach is to derive clinical relevance distributionally, i.e. to justify a value based on a failure representing a certain position in the endpoint distribution.

A key advantage of distributional failure values is that they closely relate to $\Delta$, which is the primary driver of higher relative variance, so tight control can be kept over it. For quantiles, assuming a parametric distribution such as a t-distribution should allow approximate conversion into $\Delta$. Their second advantage is that the concept is directly portable between certain trials, clinical programmes, endpoints or even indications. Although the numerical failure value and the distribution it is taken from will vary, the summary measure it is based on could remain consistent. Such standards would help with patients' and practitioners' communities' acceptance and allow easier comparison of results through meta-analysis. As noted previously though, more work needs to be done to quantify the distributional uncertainties around these approaches and appropriately account for them in estimation.

Specific reflections apply to trials that define multiple estimands, when the endpoint is measured at different timepoints. This applies both to estimands formally declared (e.g. as secondary estimands for longer-term characterisation of the effect) and to estimands implicitly defined in tables or visualisations of longitudinal data. A failure value is usually needed at each visit. Applying the same relative distributional definition at each timepoint will result in a different absolute value whenever the outcome distribution changes over time. Conversely, applying the same absolute value at every timepoint (or equivalently the same distributional definition based on the primary estimand's timepoint) will result in different relative distributional definitions, and in different relative variances at each timepoint. In many cases, it can be argued that a failure, such as participant death, is of the same severity no matter when it occurs and that therefore the same absolute failure value (or equivalently the exact same distributional definition on the primary estimand's timepoint) should be incorporated in all estimands. Whichever approach is adopted, the choice between a constant absolute value and a constant distributional definition should be made explicit when specifying failure values across multiple timepoints and when interpreting the subsequent results.

Choice of failure value should be discussed openly while designing clinical trials. As with all discussions about estimands, multi-disciplinary input is required to make appropriate decisions. Medical and statistical perspectives should be particularly sought on appropriate values, balancing clinical relevance against statistical operating characteristics. For late phase trials, it is also necessary to discuss definitions with appropriate regulatory agencies. Whatever definition is eventually selected for the failure value, it is important to clearly define it in the estimand definition in protocols, papers and the clinical trial report.

Where there is not clearcut clinical justification for the chosen failure value, it is important to perform supplementary analysis to look at the impact of alternative failure value definitions in the estimand. If there are fewer ICEs in the investigational arm compared to the control arm, increasingly penalising failure values will increase point estimates but may still reduce statistical significance. Such situations should not be interpreted as a lack of robustness.

An alternative response to higher relative variance or finding a consensus on an appropriate failure value is to replace the mean with an alternative summary measure such as the median. These alternatives are inherently more robust to extreme failure values and therefore avoid the high relative variance induced by highly penalising choices. This makes them attractive in settings where clinically appropriate failure values are necessarily extreme. However, their estimation may still have somewhat reduced efficiency compared to parametric models for means under a Normal distribution. Their estimation methods, covariate adjustment and interpretation are also less familiar to practitioners and regulators. As such, while alternative summary measures can provide a pragmatic solution, their use represents a substantive change to the estimand, introduces their own interpretation and estimation issues, and is a break from convention.

The interpretation of continuous composite endpoints will change depending on the frequency of the ICEs and extremity of the failure value. When composite-handled ICEs are rare, the interpretation will focus primarily on the continuous component and the estimates may be somewhat relatable to those from alternative ICE-handling strategies. However, as the composite-handled ICEs get more frequent, or if the failure value is extreme, the meaning of the endpoint becomes more a mixture of the continuous and 'failure' clinical concepts, similar to a utility function. In cases where the ICE and its failure value predominate, e.g. in situations with considerable use of rescue therapy, the values of the continuous component may become less relevant and the endpoint becomes more a pseudo-binary measure of the frequency of ICEs. This provides a radically different interpretation than if, for instance, a hypothetical handling approach that focussed on the values themselves had been adopted. In these cases, interpretation may be aided by moving to a simpler dichotomised endpoint. Expectations around the frequency of the relevant ICE(s) and their impact on interpretation should therefore be considered when planning such a continuous composite endpoint.

Sometimes, continuous composites may not be clinically appropriate even for ICEs such as death, notably where the relevant ICE and the continuous outcomes are only weakly clinically related over the duration of the trial. For instance, it may be believed that the ICE would not be caused by a failure of treatment. Then there may not be a meaningful single clinical concept combining both the ICE and the continuous outcome of interest. In these situations, it becomes more clinically relevant to want to know the continuous endpoint had the ICE not occurred and it also becomes difficult to clinically justify a reasonable failure value. To demonstrate this, consider the weight loss setting. Here, it might be reasonable to handle use of alternative weight-loss therapy due to a lack of efficacy as a composite, since it represents a failure of treatment. In contrast, death during the duration of the clinical trial may not be linked to a failure to lose weight and may be observed to be more associated with weight loss for other medical reasons (i.e. cachexia). The hypothetical of what the weight would have been had the participant not died might be of more clinical relevance in assessing the effectiveness of a treatment for weight loss.

## Conclusion

Continuous composite endpoints provide a coherent way to incorporate clinically meaningful intercurrent events, such as death, directly into the endpoint definition, ensuring that the occurrence of the event itself is reflected when post-event outcomes are non-existent or no longer meaningful. These endpoints retain a continuous measure of treatment effect. However, when the estimand is defined using a mean summary measure, the choice of failure value becomes a critical and consequential design decision. We derived a formula showing that the relative variance of these endpoints depends on the frequency of the ICEs and in particular the extremity of the failure value. Highly penalising values, although clinically intuitive, can substantially increase variance and reduce power. In extreme cases these values cause the endpoint to collapse into a pseudo-binary measure, that is difficult to interpret clinically.

This paper demonstrates that the failure value should not be treated as a secondary implementation detail, but as a core component of the estimand that requires explicit clinical justification and statistical evaluation. Distribution-based failure value definitions such as a quantile from a control-arm population distribution offer a pragmatic option in situations where there is no clear alternative, generating some clinical interpretability while allowing control over statistical operating characteristics. Where no acceptable failure value can balance the above considerations, alternative summary measures such as the median should be considered. Further research is needed on both the estimand definition (including multidisciplinary consensus-building) and its estimation.

## Acknowledgements

JB's contributions came from work that has been sponsored and initiated by Boehringer-Ingelheim Pharma GmbH & Co. KG. The authors are grateful to two anonymous internal reviewers for their insightful feedback during preparation of the manuscript.

## Conflict of Interest Statement

JB is a contractor at Elderbrook Solutions GmbH and his contributions came from work that has been sponsored and initiated by Boehringer-Ingelheim Pharma GmbH & Co. KG. ONK is an independent consultant to the pharmaceutical industry and to not-for-profit organisations. All other authors are employed by their respective affiliations. TD owns shares of GlaxoSmithKline. TM owns shares of Novartis Pharma AG. CBP owns shares of Novo Nordisk A/S. DW owns stock in AstraZeneca, MM is a Biogen employee and holds stock, stock options, or both, in Biogen. The views expressed in this article are the personal views of the authors and may not be understood or quoted as being made on behalf of or reflecting the position of the regulatory agencies or organizations with which the authors are affiliated.

## References

1. ICH. E9(R1) (2019) Addendum on Estimands and Sensitivity Analysis in Clinical Trials. https://www.ich.org/page/efficacy-guidelines; 2019.
2. ICH. E9(R1) Training Material. https://database.ich.org/sites/default/files/E9%28R1%29%20Training%20Material%20PDF_0.pdf; 2020.
3. Pham TM, Kahan BC, Lopes A, Rashid M, Hoskin PJ, White IR. Estimating treatment effects in trials with outcome data truncated by death: A case study on aligning estimators with estimands. Clinical Trials. 2025 Dec;22(6):676-86.
4. Keene O, Lynggaard H, Fletcher C, Wright D, Terrill P. Different Clinical Questions Need Different Estimands. Therapeutic Innovation & Regulatory Science. 2026 Jul 7:1-4.
5. Keene ON. Strategies for composite estimands in confirmatory clinical trials: examples from trials in nasal polyps and steroid reduction. Pharmaceutical Statistics. 2019 Jan;18(1):78-84.
6. Permutt T, Li F. Trimmed means for symptom trials with dropouts. Pharmaceutical statistics. 2017 Jan;16(1):20-8.
7. Kahan BC, Pham TM, Tweed C, Morris TP. When do composite estimands answer non-causal questions? *arXiv*. June 2025. doi:10.48550/arxiv.2506.22610
8. Altman DG. Practical statistics for medical research. Chapman and Hall/CRC; 1990 Nov 22.
9. European Medicines Agency. Guideline on the clinical investigation of medicines for the treatment of Alzheimer's disease. CPMP/EWP/553/95 Rev.2; 2018. Available from: https://www.ema.europa.eu/en/documents/scientific-guideline/guideline-clinical-investigation-medicines-treatment-alzheimers-disease-revision-2_en.pdf
10. European Medicines Agency. Guideline on the development of new medicinal products for the treatment of ulcerative colitis. CHMP/EWP/18463/2006 Rev.1; 2018. Available from: https://www.ema.europa.eu/en/documents/scientific-guideline/guideline-development-new-medicinal-products-treatment-ulcerative-colitis-revision-1_en.pdf
11. European Medicines Agency. Draft guideline on clinical development of medicinal products for the treatment and prevention of bipolar disorder. EMA/CHMP/406037/2024 Rev. 1 [Internet]. Amsterdam (The Netherlands): European Medicines Agency; 2024. Available from: https://www.ema.europa.eu/en/documents/scientific-guideline/draft-guideline-clinical-development-medicinal-products-treatment-prevention-bipolar-disorder-revision-1_en.pdf
12. Food and Drug Administration (FDA), U.S. Department of Health and Human Services. Amyotrophic Lateral Sclerosis: Developing Drugs for Treatment Guidance for Industry 2019. Available at: https://www.fda.gov/media/130964/download
13. Darken P, Nyberg J, Ballal S, Wright D. The attributable estimand: a new approach to account for intercurrent events. Pharmaceutical Statistics. 2020 Sep;19(5):626-35.
14. Han JK, Bachert C, Fokkens W, Desrosiers M, Wagenmann M, Lee SE, Smith SG, Martin N, Mayer B, Yancey SW, Sousa AR. Mepolizumab for chronic rhinosinusitis with nasal polyps (SYNAPSE): a randomised, double-blind, placebo-controlled, phase 3 trial. The Lancet Respiratory Medicine. 2021 Oct 1;9(10):1141-53.

15. Kahan BC, Hindley J, Edwards M, Cro S, Morris TP. The estimands framework: a primer on the ICH E9 (R1) addendum. bmj. 2024 Jan 23;384.
16. Van Lancker K, Tarima S, Bartlett J, Bauer M, Bharani-Dharan B, Bretz F, Flournoy N, Michiels H, Olarte Parra C, Rosenberger JL, Cro S. Estimands and their estimators for clinical trials impacted by the COVID-19 pandemic: a report from the NISS Ingram Olkin Forum Series on unplanned clinical trial disruptions. Statistics in Biopharmaceutical Research. 2023 Jan 2;15(1):94-111.
17. Mallinckrodt CH, Bell J, Liu G, Ratitch B, O'kelly M, Lipkovich I, Singh P, Xu L, Molenberghs G. Aligning estimators with estimands in clinical trials: putting the ICH E9 (R1) guidelines into practice. Therapeutic innovation & regulatory science. 2020 Mar 1:2168479019836979.
18. Richeldi L, Azuma A, Cottin V, et al. Nerandomilast in Patients with Idiopathic Pulmonary Fibrosis. *New England Journal of Medicine*. Published online May 18, 2025. doi:https://doi.org/10.1056/nejmoa2414108
19. Hodges JL, Lehmann EL. Estimates of Location Based on Rank Tests. *The Annals of Mathematical Statistics*. 1963;34(2):598-611. doi:https://doi.org/10.1214/aoms/1177704172
20. Nathan SD, Smith P, Deng C, et al. Inhaled treprostinil for idiopathic pulmonary fibrosis. *New England Journal of Medicine*. March 2026. doi:10.1056/nejmoa2512911
21. Mütze T, Bell J, Englert S, Hougaard P, Jackson D, Lanius V, Ravn H. Principles for defining estimands in clinical trials—A proposal. Pharmaceutical Statistics. 2025 Jan;24(1):e2432.
22. Ranu H, Wilde M, Madden B. Pulmonary function tests. *The Ulster Medical Journal*. 2011;80(2):84-90.
23. Althunian TA, de Boer A, Groenwold RHH, Klungel OH. Defining the noninferiority margin and analysing noninferiority: An overview. *British Journal of Clinical Pharmacology*. 2017;83(8):1636-1642. doi:https://doi.org/10.1111/bcp.13280
24. Klukowska AM, W. Peter Vandertop, Schröder ML, Staartjes VE. Calculation of the minimum clinically important difference (MCID) using different methodologies: case study and practical guide. *European Spine Journal*. Published online June 28, 2024. doi:https://doi.org/10.1007/s00586-024-08369-5
25. American Thoracic Society. ATS Statement: Guidelines for the Six-Minute Walk Test. *American Journal of Respiratory and Critical Care Medicine*. 2002;166(1):111-117. doi:https://doi.org/10.1164/ajrccm.166.1.at1102
26. Schmetterer L, Scholl H, Garhöfer G, et al. Endpoints for clinical trials in ophthalmology. *Progress in Retinal and Eye Research*. Published online January 2, 2023:101160. doi:https://doi.org/10.1016/j.preteyeres.2022.101160
27. Kwon SY, Kim G, Lee J, et al. Association of body weight change with all-cause and cause-specific mortality: A nationwide population-based study. *Diabetes Research and Clinical Practice*. 2023;199:110666. doi:https://doi.org/10.1016/j.diabres.2023.110666
28. Cohen J. Statistical Power Analysis for the Behavioral Sciences. 2. Hillsdale (NJ): Eribaum; 1988.
29. CLSI (Clinical and Laboratory Standards Institute). *EP28-A3c: Defining, Establishing, and Verifying Reference Intervals in the Clinical Laboratory* (3rd edition, 2010). https://clsi.org/shop/standards/ep28/

30. Cole TJ, Green PJ. Smoothing reference centile curves: The lms method and penalized likelihood. *Statistics in Medicine*. 1992;11(10):1305-1319. doi:10.1002/sim.4780111005
31. Mallinckrodt CH, Lane PW, Schnell D, Peng Y, Mancuso JP. Recommendations for the primary analysis of continuous endpoints in longitudinal clinical trials. *Drug Information Journal*. 2008;42(4):303-319. doi:10.1177/009
32. D. B. Rubin, *Multiple Imputation for Nonresponse in Surveys* (New York: John Wiley & Sons, 1987)
33. Bell J, Drury T, Mütze T, et al. Estimation Methods for Estimands Using the Treatment Policy Strategy; a Simulation Study Based on the PIONEER 1 Trial. *Pharmaceutical Statistics*. 2025;24(2). doi:https://doi.org/10.1002/pst.2472
34. Senn S. Disappointing dichotomies. Pharmaceutical statistics. 2003 Oct 1;2(4):239-40.
35. Park C, Kim H, Wang M. Investigation of finite-sample properties of robust location and scale estimators. *Communications in Statistics - Simulation and Computation*. 2020;51(5):2619-2645. doi:10.1080/03610918.2019.1699114
36. Staffa SJ, Zurakowski D. Calculation of confidence intervals for differences in medians between groups and comparison of methods. Anesthesia & Analgesia. 2020 Feb 1;130(2):542-6.
37. Carpenter J, Bithell J. Bootstrap confidence intervals: when, which, what? A practical guide for medical statisticians. Statistics in medicine. 2000 May 15;19(9):1141-64.
38. Brunner E, Vandemeulebroecke M, Mütze T. Win odds: An adaptation of the win ratio to include ties. *Statistics in Medicine*. 2021;40(14):3367-3384. doi:https://doi.org/10.1002/sim.8967

## Appendix 1: Derivation of the Relative Variance of a Composite Continuous Endpoint

Let $Y^{\kappa}$ follow a mixture distribution, where the first component has mean $\mu$ and variance $\sigma^2$ with proportion $1-p$, the second component has mean $\kappa$ and variance $0$, and proportion $p$. The well-known formula for the variance of a mixture distribution then gives:

$$VAR(Y^{\kappa}) = (1-p)\cdot\{\sigma^2+\mu^2\} + p\cdot\{0+\kappa^2\} - E(Y^{\kappa})^2. \qquad (A1)$$

Plugging in $E(Y^{\kappa}) = (1-p)\cdot\mu + p\cdot\kappa$ into (A1) then furnishes:

$$\begin{aligned} VAR(Y^{\kappa}) &= (1-p)\cdot\{\sigma^2+\mu^2\} + p\cdot\kappa^2 - (1-p)^2\cdot\mu^2 - p^2\cdot\kappa^2 + 2\cdot p\cdot(1-p)\cdot\mu\cdot\kappa \\ &= (1-p)\cdot\sigma^2 + p\cdot(1-p)\cdot\{\mu^2+\kappa^2-2\cdot\mu\cdot\kappa\} \\ &= (1-p)\cdot\sigma^2 + p\cdot(1-p)\cdot(\mu-\kappa)^2 \\ &= (1-p)\cdot\{\sigma^2 + p\cdot(\mu-\kappa)^2\}. \qquad (A2) \end{aligned}$$

The relative variance, $\varphi$ , the variance of the mixture distribution divided by the variance of the first component is then given by:

$$\varphi = \frac{VAR(Y^{\kappa})}{\sigma^2} = (1-p)\cdot\{1+p\cdot\Delta^2\},$$

where $\Delta$ is the standardised difference of the failure value from the mean of the first component:

$$\Delta = \frac{\kappa-\mu}{\sigma}$$